\documentclass[twocolumn,showpacs,aps,pra,amsmath,amssymb,superscriptaddress]{revtex4-2}

\usepackage{graphicx}% Include figure files
\usepackage{epstopdf} % Allow automatic EPS to PDF conversion
\usepackage{dcolumn}% Align table columns on decimal point
\usepackage{bm,bbm}% bold math
\usepackage{hyperref}% add hypertext capabilities
\usepackage{subfig}
\usepackage[ruled,linesnumbered]{algorithm2e}
\usepackage{soul,color,xcolor}  
\usepackage{overpic}
\usepackage{multirow}
\usepackage{float}
\usepackage{dblfloatfix}
\usepackage{placeins}

\newcommand{\beq}{\begin{equation}}
	\newcommand{\eeq}{\end{equation}}
\newcommand{\bqa}{\begin{eqnarray}}
	\newcommand{\eqa}{\end{eqnarray}}

\begin{document}
	
	\title{Resource-efficient quantum approximate optimization algorithm via Bayesian optimization and maximum-probability evaluation}
	
	\author{Siran Zhang}
	\affiliation{National Key Laboratory of Autonomous Intelligent Unmanned Systems, Shanghai Research Institute for Intelligent Autonomous Systems, Tongji University, Shanghai 201203, China}

	\author{Shuming Cheng}
	\email{drshuming.cheng@gmail.com}
		\affiliation{National Key Laboratory of Autonomous Intelligent Unmanned Systems, Shanghai Research Institute for Intelligent Autonomous Systems, Tongji University, Shanghai 201203, China}
	\affiliation{College of Electronic and Information Engineering, Tongji University, Shanghai 201804, China}%
	
	\date{\today}
	
	\begin{abstract}
The quantum approximate optimization algorithm (QAOA) is a leading variational approach to combinatorial optimization, but its practical performance depends strongly on objective design, parameter search, and shot allocation. We present a resource-efficient QAOA framework that uses the cut value of the most probable measured bitstring as the optimization objective, combines it with Bayesian optimization, and adaptively allocates shots using dual criteria based on mode confidence and normalized cut-value variance. Numerical experiments on 3-regular MaxCut show that, for both unweighted and weighted instances, the proposed scheme achieves discrete-solution quality comparable to that of the conventional expectation-based objective while typically requiring fewer total shots to reach the same final mode accuracy. These results indicate that reorganizing QAOA around the maximum-probability bitstring provides an effective route to improving practical performance under limited measurement budgets.
\end{abstract}
	
	\maketitle

\section{Introduction}\label{sec:introduction}

The Quantum Approximate Optimization Algorithm (QAOA) is one of the most representative hybrid quantum--classical algorithms for combinatorial optimization problems\cite{farhi2014quantum}. By generating a probability distribution over candidate solutions through a parameterized quantum circuit and iteratively updating the circuit parameters with a classical optimizer, QAOA provides a practically feasible route for discrete optimization tasks such as Max-Cut under shallow-circuit, noisy-hardware, and limited-resource conditions. In recent years, research on QAOA has expanded from its original algorithmic formulation to performance analysis, problem adaptation, noise robustness, hardware constraints, and various extended variants, forming a systematic and rapidly developing research landscape\cite{blekos2024review}. Consequently, how to further improve the solution quality and resource efficiency of QAOA under noisy intermediate-scale quantum (NISQ) conditions has become an important topic in this field.

For variational combinatorial optimization algorithms such as QAOA, the final performance is not determined solely by the ansatz itself, but rather by the interplay of three tightly coupled factors: how the objective function is defined, how the parameters are optimized, and how the finite number of measurement shots is allocated\cite{tilly2022variational}. Existing works usually adopt the expectation value of the problem Hamiltonian as the optimization objective. This choice is natural from the perspective of quantum mechanics, since it directly corresponds to a physical observable, and it is also convenient for theoretical analysis and classical post-processing. However, for classical combinatorial optimization tasks such as Max-Cut, the underlying problem is typically computationally hard; moreover, previous studies have pointed out that training variational quantum algorithms can itself be difficult in general settings\cite{bittel2021training}. More importantly, under a limited measurement budget, the estimation error of the objective function quickly becomes a central bottleneck to algorithmic efficiency. As a result, shot assignment should no longer be regarded as a secondary implementation detail, but rather as a key design dimension of the method itself\cite{zhu2024optimizing}. This implies that local improvements applied only to the ansatz or to a single classical optimizer are often insufficient to fundamentally improve the overall resource efficiency of QAOA.

From the perspective of task alignment, the conventional expectation-based objective suffers from a more fundamental limitation in classical discrete optimization: it is not necessarily well aligned with the form of the final output. For Max-Cut and related problems, the algorithm is ultimately required to output a concrete bitstring rather than an abstract distributional average. In other words, the traditional expectation-based objective optimizes the average quality of the sampled distribution, but does not directly target the quality of the \emph{most likely} solution to be measured. This mismatch between the training objective and the task output becomes even more pronounced under a limited measurement budget. On the one hand, obtaining a stable estimate of the expectation value usually requires evaluating multiple sampled bitstrings and aggregating them with appropriate weights; on the other hand, what the actual optimization task cares about is whether the algorithm can output a high-quality discrete solution with sufficiently high probability. Existing studies have also shown that QAOA is inherently a sampling-based algorithm, and its performance can be characterized by the approximation quality of individual samples, or by the probability of observing a solution above a target threshold, rather than solely by the mean value of the distribution\cite{Larkin_2022}. Therefore, rethinking the objective function of QAOA in classical combinatorial optimization is a natural starting point for improving both task alignment and resource efficiency.

In addition to the objective design, the choice of parameter optimization strategy also has a profound impact on the overall resource consumption of QAOA. Previous studies have systematically compared different classical optimizers for QAOA in solving Max-Cut and found that different optimizers may lead to substantial differences in convergence speed, the number of objective evaluations, and the final solution quality\cite{Fern_ndez_Pend_s_2022}. For QAOA executed on near-term quantum devices, each parameter evaluation typically requires actual circuit execution and measurement, which makes the optimization process essentially a black-box problem characterized by expensive evaluations, a limited number of samples, and noisy observations. In this regime, Bayesian optimization (BO) is particularly appealing because it can exploit historical evaluation data to construct a surrogate model and efficiently select new query points under a limited budget\cite{Jones_1998,NIPS2012_0531165}. Furthermore, QAOA-oriented studies have shown that BO can effectively reduce the number of quantum circuit calls and exhibits favorable robustness in low-depth and noisy settings\cite{BFQAOA10286414}. Therefore, once a new objective function is introduced, it is equally important to equip it with a compatible and resource-efficient parameter search mechanism.

At the same time, we argue that under finite-shot conditions, the training of QAOA should be understood as an intrinsically stochastic optimization process rather than an idealized deterministic continuous optimization problem. Existing theoretical work has shown that, for representative variational quantum algorithms such as QAOA and VQE, objective and gradient estimation under finite measurements naturally fall into the framework of stochastic optimization; accordingly, variational optimization is more appropriately interpreted as finite-shot stochastic optimization rather than exact deterministic optimization\cite{Sweke_2020}. This perspective leads to a direct and important implication: shots should not be treated as a fixed external budget, but as a dynamically adjustable resource within the optimization process itself. Further studies have also indicated that adaptive shot allocation can not only reduce the total measurement cost, but may also improve the convergence behavior of the optimization procedure\cite{Adaptive}. Therefore, under an overall limited resource budget, how to decide whether additional measurements are necessary according to the statistical uncertainty at the current parameter point, and how to preferentially allocate the finite measurement budget to the most promising candidate regions, become central questions addressed in this work.

Motivated by the above observations, this paper does not view QAOA training simply as parameter tuning under a fixed objective. Instead, we formulate it as a resource-efficiency problem under a limited quantum sampling budget, where the objective function, the classical optimizer, and the sampling mechanism should be designed jointly. Specifically, we replace the conventional expectation value with the objective value associated with the maximum-probability bitstring, so that the optimization target is more directly aligned with the final discrete output. We employ Bayesian optimization to handle the high noise, expensive evaluation, and potential nonsmoothness introduced by this new objective. We further combine \texttt{mode\_confidence} and sample variance to construct an adaptive sampling mechanism: for each candidate parameter point, a small number of shots is first used to quickly assess whether the current mode is sufficiently stable, and additional shots are allocated only when warranted by statistical uncertainty. At the same time, an early-stopping mechanism is introduced to terminate redundant evaluations when the marginal gain in the later stage of optimization becomes negligible, thereby reducing unnecessary quantum resource consumption. We also consider an optional engineering enhancement module that can further improve the output stability of the target solution after a promising solution has been identified in Stage 1, highlighting the practical extensibility of the proposed framework.

The main contributions of this work can be summarized as follows:

1. We propose a mode-based objective that shifts the optimization target of QAOA from the distributional mean to the quality of the most likely output solution, thereby improving the alignment between the training objective and the output form of discrete optimization tasks.

2. We develop a unified framework that integrates Bayesian optimization, mode-based evaluation, adaptive sampling, and early stopping, enabling objective design, parameter search, and measurement resource allocation to work cooperatively under a finite-shot setting.

3. Beyond the main Stage-1 framework, we discuss an optional engineering refinement to improve the output stability of the target solution after a promising solution has been found, highlighting the practical deployment potential of the proposed approach.

Compared with existing works that mainly focus on expectation-based objectives, a single classical optimizer, or isolated shot-allocation strategies, this paper emphasizes a resource-efficiency-oriented co-design paradigm. Rather than treating the objective function, optimizer, and sampling budget as separate components, we jointly optimize them around a central goal: obtaining the highest-quality discrete solution possible with the least possible quantum resources.

\section{Theoretical Foundations and Methodology}
\label{sec:theory_method}

\subsection{Problem Formulation}
\label{subsec:problem_overview}

In this work, we take the (possibly weighted) MaxCut problem as a representative task in discrete combinatorial optimization. Given an undirected weighted graph
$G=(V,E,w)$, where $|V|=n$, the objective is to maximize the cut value over a binary partition
$z=(z_1,\ldots,z_n)\in\{0,1\}^n$:
Here, $V$ denotes the vertex set, $E\subseteq V\times V$ denotes the edge set, and $w:E\to\mathbb{R}\ge 0$ assigns edge weights $w_{uv}$ to each $(u,v)\in E$. The binary variable $z_i\in\{0,1\}$ indicates the side of vertex $i$ in the cut, and $\mathbf{1}(\cdot)$ is the indicator function that equals $1$ when its argument is true and $0$ otherwise.
\begin{equation}
C(z)=\sum_{(u,v)\in E} w_{uv}\,\mathbf{1}(z_u\neq z_v).
\label{eq:maxcut_cut}
\end{equation}
The corresponding cost Hamiltonian can be written as
\begin{equation}
H_C=\frac{1}{2}\sum_{(u,v)\in E} w_{uv}\left(I-Z_uZ_v\right),
\label{eq:maxcut_cost_hamiltonian}
\end{equation}
which is diagonal in the computational basis. Consequently, a measured bit string can be mapped directly to a discrete solution and its associated cut value.

QAOA may be viewed as a finite-depth discrete approximation to adiabatic evolution. Introducing the mixer Hamiltonian
\begin{equation}
H_M=\sum_{i=1}^{n}X_i,
\label{eq:mixer_hamiltonian}
\end{equation}
and choosing $|+\rangle^{\otimes n}$ as the initial state, the depth-$p$ QAOA ansatz is
\begin{equation}
|\psi(\bm{\beta},\bm{\gamma})\rangle=
\prod_{\ell=1}^{p}
e^{-i\beta_\ell H_M}
e^{-i\gamma_\ell H_C}
|+\rangle^{\otimes n},
\label{eq:qaoa_state}
\end{equation}
where $\bm{\beta}=(\beta_1,\ldots,\beta_p)$,
$\bm{\gamma}=(\gamma_1,\ldots,\gamma_p)$, and
$\bm{\theta}=[\bm{\beta},\bm{\gamma}]\in\mathbb{R}^{2p}$.
Within a classical--quantum hybrid optimization loop, one repeatedly evaluates the quality of the quantum state associated with a given parameter vector $\bm{\theta}$ and updates the parameters accordingly.

The conventional approach typically uses the expectation value of the cost Hamiltonian as the objective function:
\begin{equation}
F_{\mathrm{exp}}(\bm{\theta})
=
\langle\psi(\bm{\theta})|H_C|\psi(\bm{\theta})\rangle
=
\sum_{z\in\{0,1\}^n}p_{\bm{\theta}}(z)\,C(z),
\label{eq:traditional_expectation}
\end{equation}
where $p_{\bm{\theta}}(z)$ denotes the probability of measuring bit string $z$. Under finite sampling, if the total number of measurements is $N$ and the empirical frequency is denoted by
$\hat p_{\bm{\theta}}(z)=n_z/N$, then the corresponding estimator is
\begin{equation}
\hat F_{\mathrm{exp}}(\bm{\theta})
=
\sum_{z}\hat p_{\bm{\theta}}(z)\,C(z)
=
\frac{1}{N}\sum_{s=1}^{N}C\!\left(z^{(s)}\right).
\label{eq:traditional_expectation_estimator}
\end{equation}
This objective is relatively smooth and is therefore convenient to combine with gradient-based or other continuous optimization methods. However, it optimizes the average performance of the output distribution rather than the quality of the single discrete solution ultimately returned.

However, the traditional expectation-value estimator is not without cost. For a given parameter point, if the total number of measurements is $N$, then the classical post-processing typically requires computing the cut value $C(z)$ for each sampled bit string at least once (or, equivalently, processing all edge-related contributions), leading to a per-point evaluation cost of approximately $O(N|E|)$. When this estimate must be repeated over a large number of parameter points during the optimization process, the total resource consumption on both the classical and quantum sides grows rapidly. In particular, under fixed high-shot evaluations combined with repeated outer-loop searches, the potential advantage of the hybrid algorithm may be substantially diluted by the high evaluation overhead. It is therefore desirable to design an objective-evaluation strategy that both economizes quantum sampling and reduces the classical post-processing burden.

For combinatorial optimization, what ultimately matters is whether the quantum circuit can output a high-quality discrete solution under a limited measurement budget, rather than merely achieving a large expectation value. In particular, when quantum sampling resources are scarce, continuing to use a large fixed number of shots at every candidate parameter point for a high-precision expectation estimate can waste substantial resources on regions of the parameter space that do not merit close examination. Motivated by this consideration, we shift the emphasis of this work from ``how to estimate the expectation value more accurately'' to ``how to find high-quality discrete solutions more efficiently under a limited quantum sampling budget.''

Accordingly, at the level of performance evaluation, we emphasize the quality of the solution most likely to be returned in the end. Let
$z_{\mathrm{mode}}^{\mathrm{final}}$ denote the most frequently observed bit string under the final high-precision evaluation, and let $z^\star$ be the exact optimal solution. We define the primary metric as
\begin{equation}
\mathrm{Acc}_{\mathrm{mode}}
=
\frac{C\!\left(z_{\mathrm{mode}}^{\mathrm{final}}\right)}{C(z^\star)}.
\label{eq:final_mode_accuracy}
\end{equation}
This definition is consistent with the quantity denoted \textit{final\_mode\_accuracy} in the code implementation and is also more closely aligned with the actual output form of quantum optimization in discrete decision tasks.

\subsection{Methodology}
\label{subsec:basic_principles}

To better align the optimization target with the final discrete output, we adopt a mode-based objective built on the most frequently observed bit string. For a given parameter point $\bm{\theta}$, let the measurement counts be $\{n_z\}$. We then define
\begin{equation}
z_{\mathrm{mode}}(\bm{\theta})
=
\arg\max_{z} n_z,
\qquad
\hat F_{\mathrm{MAP}}(\bm{\theta})
=
C\!\left(z_{\mathrm{mode}}(\bm{\theta})\right).
\label{eq:map_objective}
\end{equation}
This corresponds to the cut value of the bit string that appears most frequently in the current measurement outcomes. Compared with the conventional objective in \eqref{eq:traditional_expectation_estimator}, this objective no longer averages over all samples. Instead, it focuses directly on the discrete solution that the quantum circuit is currently most likely to output, thereby making the optimization target more consistent with the practical goal of returning a high-quality feasible solution in a combinatorial optimization task.

The mode-based objective brings two immediate changes. First, it places ``what the circuit is most likely to output'' rather than ``how good the distribution is on average'' at the center of the evaluation, and is therefore better aligned with the final goal of a discrete decision problem. Second, it no longer requires a uniformly precise estimate of the entire sampling distribution; instead, it prioritizes determining whether the dominant output solution has already become sufficiently clear. Under a limited quantum sampling budget, this shift allows evaluation resources to be released from a large number of noncritical samples and redirected toward parameter regions that genuinely require further discrimination.

At the same time, however, this objective also makes the optimization landscape more discrete. Because the objective value is determined by the modal bit string, as long as the dominant bit string does not change,
\begin{equation}
z_{\mathrm{mode}}(\bm{\theta}+\Delta\bm{\theta})
=
z_{\mathrm{mode}}(\bm{\theta})
\;\Longrightarrow\;
\hat F_{\mathrm{MAP}}(\bm{\theta}+\Delta\bm{\theta})
=
\hat F_{\mathrm{MAP}}(\bm{\theta}),
\label{eq:flat_map_landscape}
\end{equation}
so the parameter-space landscape becomes distinctly piecewise flat, with weak or even vanishing gradients. For this reason, conventional gradient-descent methods that rely on local smoothness often fail to produce stable updates, and we therefore adopt Bayesian optimization as the primary search strategy in Stage-1, where the objective is black-box, discrete, and noisy.

Specifically, rather than assigning a large fixed number of shots to every candidate parameter point from the outset, we first perform a small-shot pilot evaluation and then decide, based on the pilot result, whether additional measurement resources should be allocated. Let the current accumulated number of shots be $N$, let the empirical distribution be $\hat{\bm{p}}$, and let the mode be $z_{\mathrm{mode}}$. To determine whether the leading bit string is already sufficiently stable, we use the mode confidence as a first criterion. In the default implementation, this quantity is estimated by bootstrap resampling:
\begin{equation}
\mathrm{Conf}_{\mathrm{mode}}
=
\mathbb{P}_{\tilde{\bm{n}}\sim\mathrm{Multinomial}(N,\hat{\bm{p}})}
\!\left[
\arg\max_{z}\tilde n_z=z_{\mathrm{mode}}
\right].
\label{eq:mode_confidence_bootstrap}
\end{equation}

Mode confidence alone is still insufficient to characterize the output stability of the current parameter point. Even if the leading bit string is temporarily ahead, if the cut-value distribution of the sampled outputs remains broad, additional sampling may still alter the judgment of which discrete solution is ultimately worth retaining. To address this issue, we further introduce the empirical variance of cut values:
\begin{equation}
\widehat{\mathrm{Var}}_{C}
=
\sum_{z}\hat p_{\bm{\theta}}(z)\,C(z)^2
-
\left(
\sum_{z}\hat p_{\bm{\theta}}(z)\,C(z)
\right)^2,
\label{eq:cut_variance}
\end{equation}
and normalize it using the graph-dependent upper bound
\begin{equation}
U_C=\sum_{(u,v)\in E}w_{uv}
\label{eq:cut_upper_bound}
\end{equation}
to obtain
\begin{equation}
\widetilde{\mathrm{Var}}_{C}
=
\frac{\widehat{\mathrm{Var}}_{C}}{U_C^2}.
\label{eq:normalized_cut_variance}
\end{equation}

Thus, rather than relying solely on the variance or solely on the mode confidence, we use a more robust dual criterion:
\begin{equation}
\mathrm{Conf}_{\mathrm{mode}}\ge \tau_{\mathrm{conf}}
\quad \text{and} \quad
\widetilde{\mathrm{Var}}_{C}\le \tau_{\mathrm{var}}.
\label{eq:dual_gate}
\end{equation}
The current evaluation at a parameter point is accepted only when both conditions are satisfied; otherwise, additional shots are allocated. If $b_r$ denotes the newly added batch size at round $r$, then the recursion can be written as
\begin{equation}
b_{r+1}
=
\min\!\left\{
\left\lfloor \rho\, b_r \right\rceil,\;
N_{\max}-\sum_{j=1}^{r}b_j
\right\},
\label{eq:adaptive_batch_growth}
\end{equation}
where $\rho>1$ is the growth factor and $N_{\max}$ is the maximum shot budget allowed for evaluating a single parameter point. Consistent with the main code path, the experiments often use $b_1=100$, $\rho=2.0$, $\tau_{\mathrm{conf}}=0.90$, $\tau_{\mathrm{var}}=0.02$, and set the per-point cap to $N_{\max}=1200$.

The key idea of this strategy is straightforward: no further sampling is wasted on parameter points that are already evidently stable, whereas for points whose output distributions remain disordered, whose modes are not yet reliable, or whose cut values still fluctuate substantially, the measurement precision is increased gradually. As a result, the quantum sampling budget is no longer distributed uniformly over all parameter points, but is instead concentrated on regions that truly require further discrimination. This stands in sharp contrast to the conventional fixed-shot expectation evaluator, which typically assigns the same large number of shots to every parameter point (for example, $1000$ shots in the code baseline). Although the latter is simple to implement and yields a smoother objective landscape, it continues to incur unnecessary measurement costs on many points that are either obviously poor or already sufficiently stable.

At the parameter-search level, we employ Bayesian optimization, which does not rely on analytic gradients, in order to match the discreteness, noise, and piecewise-flat structure exhibited by the mode-based objective under finite shots. Let the data observed up to iteration $t$ be
\begin{equation}
\mathcal{D}_t=\left\{(\bm{\theta}_i,y_i)\right\}_{i=1}^{t},
\qquad
y_i=\hat F_{\mathrm{MAP}}(\bm{\theta}_i),
\label{eq:bo_dataset}
\end{equation}
Bayesian optimization then constructs a surrogate model from the historical samples and selects new candidate parameter points by balancing exploration and exploitation. Compared with gradient-based methods, it does not require differentiability or local smoothness of the objective. Compared with exhaustive or purely random search, it can exploit historical observations to focus evaluations more efficiently on potentially high-quality regions. In the practical implementation, we use Optuna's TPE sampler as the BO backend to perform sequential search over $\bm{\theta}\in\mathbb{R}^{2p}$, together with an early-stopping mechanism based on stagnation detection so as to avoid further expenditure of trial budget and measurement resources once improvement has clearly plateaued.

\subsection{Workflow and Resource Estimation}
\label{subsec:workflow}

The Stage-1 workflow follows a simple resource-aware principle: Bayesian optimization proposes where to evaluate, while the adaptive-shot module decides how much quantum resource should be spent at that parameter point. More precisely, each candidate parameter vector is first subjected to a small-shot pilot evaluation. The resulting samples are then used to compute both the current mode-based objective and the two stability indicators in \eqref{eq:dual_gate}. Only when the mode is sufficiently dominant and the normalized cut variance is already small do we accept the current evaluation immediately; otherwise, additional shots are allocated gradually until the dual criterion is met or the per-point budget cap is reached.

This workflow reduces quantum-resource consumption in two complementary ways. First, at the \emph{objective-evaluation level}, it avoids assigning the same large shot count to all parameter points, so evidently stable or evidently poor regions can be filtered using only a modest pilot budget. Second, at the \emph{parameter-search level}, Bayesian optimization uses the accumulated history to focus future evaluations on more promising regions, thereby reducing the total number of expensive circuit evaluations required to approach a high-quality discrete solution. The complete Stage-1 procedure is summarized in Algorithm~\ref{alg:map_bo_workflow}.

\begin{algorithm}[t]
\caption{Stage-1 workflow: BO-guided mode-based QAOA with adaptive shots}
\label{alg:map_bo_workflow}
\KwIn{MaxCut instance $G$, QAOA depth $p$, initial pilot shots $b_1$, growth factor $\rho$, per-point shot cap $N_{\max}$, thresholds $\tau_{\mathrm{conf}}$ and $\tau_{\mathrm{var}}$, BO trial budget $T_{\max}$}
Initialize Bayesian optimizer and history $\mathcal{D}_0\leftarrow\emptyset$\;
\For{$t\leftarrow 1$ \KwTo $T_{\max}$}{
    Propose a candidate parameter vector $\bm{\theta}_t$ via BO\;
    $N\leftarrow 0$, $b\leftarrow b_1$\;
    \While{$N < N_{\max}$}{
        Run the QAOA circuit at $\bm{\theta}_t$ with $b$ additional shots\;
        Update counts $\{n_z\}$ and total shots $N\leftarrow N+b$\;
        Compute $z_{\mathrm{mode}}\leftarrow \arg\max_z n_z$ and $y_t\leftarrow C(z_{\mathrm{mode}})$\;
        Estimate $\mathrm{Conf}_{\mathrm{mode}}$ and $\widetilde{\mathrm{Var}}_C$\;
        \If{$\mathrm{Conf}_{\mathrm{mode}}\ge \tau_{\mathrm{conf}}$ \textbf{and} $\widetilde{\mathrm{Var}}_C\le \tau_{\mathrm{var}}$}{
            \textbf{accept} current evaluation\;
            \textbf{break}\;
        }
        $b\leftarrow \min\!\left\{\lfloor \rho b \rceil,\, N_{\max}-N\right\}$\;
    }
    Add $(\bm{\theta}_t,y_t)$ to $\mathcal{D}_t$ and update the BO surrogate\;
    Update the best-so-far solution if $y_t$ improves the incumbent\;
    \If{BO stagnation criterion is triggered}{
        \textbf{stop}\;
    }
}
\Return best parameter vector and best discrete solution found in Stage-1\;
\end{algorithm}

In the current implementation, Stage-2 is only an optional engineering refinement and is not the main methodological contribution. After Stage-1 finishes, the algorithm takes the best discrete solution observed during the search, denoted by $z_{\mathrm{tar}}$, and locally increases its sampling probability via
\begin{equation}
F_{\mathrm{amp}}(\bm{\theta})
=
p_{\bm{\theta}}(z_{\mathrm{tar}}).
\label{eq:stage2_probability_objective}
\end{equation}
Rather than optimizing the full expectation again, this step simply aims to make the already identified high-quality bit string easier to observe in the final measurement.

To control the extra cost, the code uses a lightweight randomized parameter-shift update: at each step, it selects only one parameter coordinate, and then samples only one associated physical gate to form a $\pm\pi/2$ shift estimate. The resulting gradient estimate is used in a stochastic optimizer (Adam by default), with periodic reevaluation of $p_{\bm{\theta}}(z_{\mathrm{tar}})$. Therefore, Stage-2 should be interpreted as a low-cost probability-amplification module for practical deployment, rather than as a second full-scale optimization stage.

In one sentence, the resource advantage comes from reducing the average quantum shots via adaptive evaluation and reducing classical post-processing from shot-wise scaling to distinct-outcome-wise scaling whenever $K_i\ll N_i$.

To compare the resource consumption of different methods more clearly, let the number of edges be $m=|E|$, the parameter dimension be $d=2p$, and the numbers of outer-loop parameter evaluations for the expectation-based and mode-based methods be $T_{\mathrm{exp}}$ and $T_{\mathrm{MAP}}$, respectively. Let the fixed-shot baseline use $N_{\mathrm{fix}}$ shots per evaluation, and let the adaptive method use $N_i$ shots for the $i$th parameter point, with average
\begin{equation}
\bar N_{\mathrm{adp}}
=
\frac{1}{T_{\mathrm{MAP}}}\sum_{i=1}^{T_{\mathrm{MAP}}}N_i.
\label{eq:avg_adaptive_shots}
\end{equation}
At the same time, let $K_i$ denote the number of distinct bit strings that actually appear in the evaluation of the $i$th parameter point, and define the average as
\begin{equation}
\bar K
=
\frac{1}{T_{\mathrm{MAP}}}\sum_{i=1}^{T_{\mathrm{MAP}}}K_i,
\qquad
K_i\le N_i.
\label{eq:avg_distinct_states}
\end{equation}

If the conventional method uses fixed-shot expectation evaluation together with Bayesian optimization or another zeroth-order search strategy, then its quantum and classical resource costs can be written as
\begin{equation}
\mathcal{R}^{\mathrm{q}}_{\mathrm{exp}}
=
T_{\mathrm{exp}}N_{\mathrm{fix}},
\qquad
\mathcal{R}^{\mathrm{cl}}_{\mathrm{exp}}
=
\Theta\!\left(T_{\mathrm{exp}}N_{\mathrm{fix}}m\right),
\label{eq:exp_resource_bo}
\end{equation}
where the classical complexity arises because, for each parameter point and each sampled bit string, one typically needs at least one cut-value evaluation (or, equivalently, one pass over the edge terms). If full parameter-shift gradient estimation is further incorporated, the resource consumption is amplified to
\begin{equation}
\begin{aligned}
\mathcal{R}^{\mathrm{q}}_{\mathrm{exp\text{-}grad}}
&=
\Theta\!\left(T_{\mathrm{exp}}(2d+1)N_{\mathrm{fix}}\right),\\
\mathcal{R}^{\mathrm{cl}}_{\mathrm{exp\text{-}grad}}
&=
\Theta\!\left(T_{\mathrm{exp}}(2d+1)N_{\mathrm{fix}}m\right),
\end{aligned}
\label{eq:exp_resource_grad}
\end{equation}
showing that once the traditional expectation-based method is coupled to a gradient-type outer loop, both quantum and classical costs increase linearly with the parameter dimension.

In contrast, the quantum resource of Stage-1 in the proposed method is determined by the adaptive shots:
\begin{equation}
\mathcal{R}^{\mathrm{q}}_{\mathrm{MAP}}
=
\sum_{i=1}^{T_{\mathrm{MAP}}}N_i
=
T_{\mathrm{MAP}}\bar N_{\mathrm{adp}},
\qquad
b_1 \le N_i \le N_{\max}.
\label{eq:map_quantum_resource}
\end{equation}
If one considers only the mode-based objective itself, then each parameter point requires only one counting pass through the samples, followed by a single cut-value evaluation once the final mode has been identified. The corresponding classical post-processing cost can therefore be written as
\begin{equation}
\mathcal{R}^{\mathrm{cl,obj}}_{\mathrm{MAP}}
=
\Theta\!\left(
\sum_{i=1}^{T_{\mathrm{MAP}}}N_i
+
T_{\mathrm{MAP}}m
\right).
\label{eq:map_classical_obj_resource}
\end{equation}
This means that, at the level of objective computation alone, the classical complexity can be reduced from the shot-wise $\Theta(Nm)$ of the conventional expectation method to $\Theta(N+m)$, consisting of one-pass counting plus a single cut-value evaluation.

If one further incorporates the dual criterion used in the current implementation, namely simultaneous computation of the empirical variance and $B$ bootstrap confidence estimates, then a more complete classical resource model is
\begin{equation}
\mathcal{R}^{\mathrm{cl,full}}_{\mathrm{MAP}}
=
\Theta\!\left(
\sum_{i=1}^{T_{\mathrm{MAP}}}N_i
+
\sum_{i=1}^{T_{\mathrm{MAP}}}K_i m
+
B\sum_{i=1}^{T_{\mathrm{MAP}}}K_i
\right).
\label{eq:map_classical_full_resource}
\end{equation}
Compared with \eqref{eq:exp_resource_bo}, the most important change is that edge scans and cut-value evaluations are no longer charged according to the total number of shots $N_i$, but according to the number of distinct bit strings actually observed, $K_i$. For parameter regions in which a clear mode has already emerged, one typically has $K_i\ll N_i$, and the classical post-processing overhead is therefore substantially reduced. Even in the worst case $K_i=N_i$, the method is still no worse in order than the conventional shot-wise evaluation.

Furthermore, the resource-saving factors of the two approaches can be expressed as
\begin{equation}
S_{\mathrm{q}}
=
\frac{\mathcal{R}^{\mathrm{q}}_{\mathrm{exp}}}{\mathcal{R}^{\mathrm{q}}_{\mathrm{MAP}}}
\approx
\frac{T_{\mathrm{exp}}N_{\mathrm{fix}}}
{T_{\mathrm{MAP}}\bar N_{\mathrm{adp}}},
\label{eq:quantum_saving_ratio}
\end{equation}
and
\begin{equation}
S_{\mathrm{cl}}
=
\frac{\mathcal{R}^{\mathrm{cl}}_{\mathrm{exp}}}
{\mathcal{R}^{\mathrm{cl,full}}_{\mathrm{MAP}}}
\approx
\frac{T_{\mathrm{exp}}N_{\mathrm{fix}}m}
{T_{\mathrm{MAP}}\bar N_{\mathrm{adp}}
+
T_{\mathrm{MAP}}\bar K m
+
BT_{\mathrm{MAP}}\bar K}.
\label{eq:classical_saving_ratio}
\end{equation}
If one first ignores differences in the numbers of trials and takes $T_{\mathrm{exp}}\approx T_{\mathrm{MAP}}$, then using the typical code setting $N_{\mathrm{fix}}=1000$ as a reference, the adaptive mechanism reduces the average number of shots to $\bar N_{\mathrm{adp}}\in[250,400]$, corresponding to a quantum-sampling reduction of roughly $60\%\sim75\%$. If many parameter points are accepted already near the pilot stage (for example, $N_i\approx100$), then under favorable conditions the ideal quantum-resource savings can approach $90\%$. On the classical side, if one considers only the mode objective itself, the per-point evaluation can drop from $\Theta(N_{\mathrm{fix}}m)$ to $\Theta(\bar N_{\mathrm{adp}}+m)$, yielding the theoretical saving factor
\begin{equation}
\widetilde S_{\mathrm{cl,obj}}
\approx
\frac{N_{\mathrm{fix}}m}{\bar N_{\mathrm{adp}}+m},
\label{eq:classical_obj_saving_ratio}
\end{equation}
which may be interpreted as an improvement of order $O(\min\{N_{\mathrm{fix}},m\})$. Once the variance and bootstrap terms are included, the actual savings also depend on the distinct-outcome ratio $K_i/N_i$ (or its average), but as soon as the sampled distribution begins to contract onto a small number of dominant bit strings, a substantial classical-side gain can still be maintained.

For completeness, if Stage-2 runs for $L$ steps and each shifted circuit uses $N_{\mathrm{amp}}$ shots, then the randomized single-coordinate update incurs a quantum overhead of order
\begin{equation}
\mathcal{R}^{\mathrm{q}}_{\mathrm{Stage2}}
=
\Theta\!\left(2L N_{\mathrm{amp}}\right),
\label{eq:stage2_resource_random}
\end{equation}
whereas a full parameter-shift update over all $d=2p$ parameters would scale as $\Theta(2dL N_{\mathrm{amp}})$. Hence, the implemented Stage-2 adds only a modest refinement cost and remains secondary to the main Stage-1 resource savings.

Taken together, the proposed framework does not by itself constitute a proof of a strict asymptotic quantum advantage for MaxCut in the complexity-theoretic sense. A more accurate statement is that, by shifting the objective from ``high-precision expectation estimation'' to ``identification of the dominant discrete solution,'' and by combining adaptive shots, Bayesian-optimization-based search, and an optional low-cost probability-amplification refinement, the present method substantially reduces both the quantum sampling cost and the classical post-processing cost of the hybrid optimization loop, thereby lowering the resource threshold at which practical quantum utility or an observable quantum advantage may emerge. In particular, relative to the fixed-shot expectation approach, this framework more readily satisfies the operating regime in which the total resource consumption remains well below that of high-precision classical enumeration or large-scale fixed-shot search, and is therefore more favorable for demonstrating tangible benefits of quantum methods under limited hardware budgets.

\section{Numerical Experiments and Results Analysis}
\label{sec:numerical_results}

\subsection{Experimental Setup and Chapter Organization}
\label{subsec:exp_setup}

This chapter is organized around three closely related questions: in discrete combinatorial optimization, how should the objective function be defined so that it more directly corresponds to the high-quality bitstrings ultimately returned by measurement; under the coexistence of finite sampling and a nonsmooth objective, what optimization strategy should be used for the outer-loop parameter search; and when the total measurement budget is limited, how should the number of shots be allocated across different parameter points. To address these questions within a unified setting, we adopt 3-regular MaxCut as the benchmark task throughout and compare the proposed method against two baselines under the same experimental framework: conventional expectation with Bayesian optimization, and conventional expectation with gradient descent.

The main experiments include two scaling studies. The first is a qubit-scaling experiment, where the QAOA depth is fixed at $p=2$ and the number of qubits varies as $n\in\{3,4,6,8,10,12\}$. The second is a circuit-depth-scaling experiment, where $n=10$ is fixed and the depth varies as $p\in\{1,2,3,4,5,6\}$. For each value on the horizontal axis, we randomly generate $10$ problem instances and report aggregated statistics, thereby evaluating the average performance of each method across different scales. To further assess robustness under noise, we also conduct a depolarizing-noise sweep. Specifically, with $n=10$ and $p=2$ fixed, the noise strength is increased from $0$ to $0.01$, and multiple random instances are evaluated at each noise level.

Regarding evaluation metrics, we use the final mode accuracy as the primary metric, which measures the quality of the solution corresponding to the most frequently observed bitstring in the final measurement distribution. We additionally report the final expectation accuracy, the final best-sample accuracy, and the total number of shots, which respectively characterize the average objective value, the best single-sample quality, and the overall resource consumption. For equal-accuracy comparisons, we further fix the target accuracy threshold at $0.80$ and record the cumulative number of shots consumed when each method first reaches this threshold. This directly quantifies how many quantum sampling resources are required to attain the same discrete-solution quality. With this setup, the resource-efficiency discussion in the following sections no longer remains at the level of average-accuracy comparison, but instead focuses on the joint trade-off between quantum sampling cost and final output quality.

The organization of this chapter follows the three questions above. Section~\ref{subsec:objective_design} first explains why final mode accuracy should be regarded as the more central evaluation criterion. Section~\ref{subsec:shot_allocation} then analyzes the resource savings enabled by adaptive shot allocation. Section~\ref{subsec:overall_results} presents the overall results after combining the three design choices and also briefly discusses the engineering-enhancement role of Stage~2. Finally, Section~\ref{subsec:noise_results} examines how the accuracy--resource trade-off changes in the presence of noise. 

\subsection{Objective Design: From Conventional Expectation to the Maximum-Probability Bitstring}
\label{subsec:objective_design}

\begin{figure*}[!tbp]
    \centering
    \includegraphics[width=0.90\textwidth]{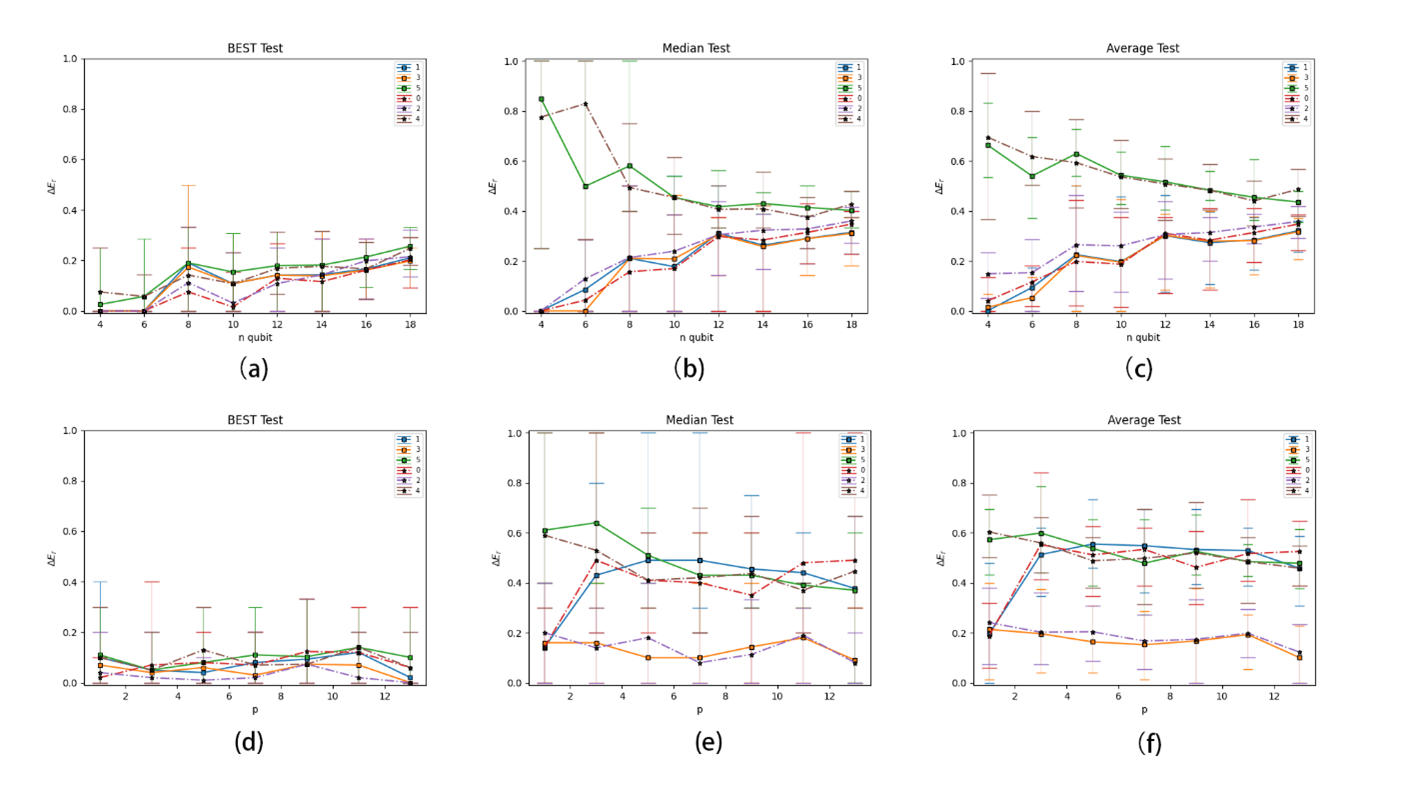}
    \caption{Comparison between the conventional expectation and the mode-based objective on unweighted 3-regular MaxCut.}
    \label{fig:pre_sameweight}
\end{figure*}

Redefining the objective function is the first question this work must answer. For 3-regular MaxCut, Fig.~\ref{fig:pre_sameweight} presents a controlled comparison between the conventional expectation and the mode-based objective in the unweighted case. As can be seen, across different optimizers and different statistical summaries, the two objectives yield broadly similar solution quality. This indicates that using the most frequently observed bitstring in the current measurement distribution as the representative output does not substantially degrade the quality of the solution itself. For the purposes of this work, this does not imply that the two objectives are fully equivalent. Rather, it provides empirical justification for using the maximum-probability-bitstring objective as a viable surrogate, which can then be further combined with new search strategies and resource-allocation policies.

\begin{figure*}[!tbp]
    \centering
    \includegraphics[width=0.90\textwidth]{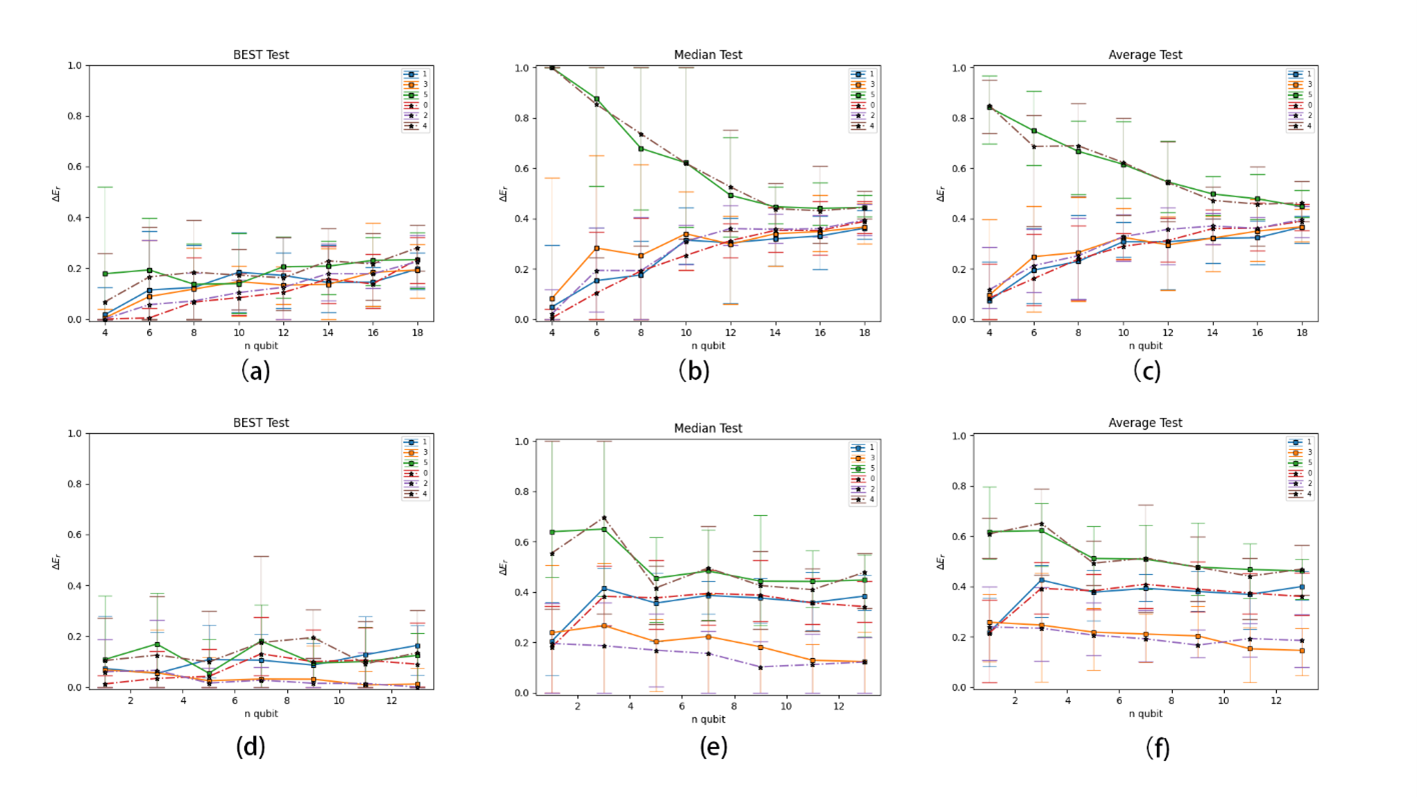}
    \caption{Comparison between the conventional expectation and the mode-based objective on weighted 3-regular MaxCut.}
    \label{fig:pre_weighted}
\end{figure*}

Figure~\ref{fig:pre_weighted} shows the corresponding results on weighted 3-regular graphs. Consistent with the unweighted setting, the mode-based objective continues to achieve solution quality close to that of the conventional expectation across different problem sizes and circuit depths. This suggests that shifting the objective from an average expectation to the maximum-probability bitstring does not rely on the idealized unit-weight assumption, but remains feasible for more general weighted MaxCut instances as well. Therefore, before moving to the main experiments, we can already draw a clear conclusion: if the ultimate goal is to output high-quality discrete solutions, then redefining the objective around the maximum-probability bitstring is both reasonable and practically operable.

 Compared with the conventional expectation, the landscape induced by the mode-based objective is more clearly discretized and flatter. On the one hand, this weakens the usefulness of local gradient information, making optimization strategies that rely purely on local smoothness less appropriate. On the other hand, it makes the objective more directly centered on the question of which bitstring will become the dominant output. For this reason, the subsequent experiments do not treat the objective function, the optimization strategy, and shot allocation as isolated design choices; instead, they are evaluated jointly within a unified framework.

Once the objective is specified, the second question is how to choose the outer-loop classical optimizer. Under finite-shot conditions, the objective behaves more like a noisy, piecewise-flat black-box optimization problem. We therefore adopt Bayesian optimization as the unified outer-loop search engine, and use conventional expectation with Bayesian optimization and conventional expectation with gradient descent as baselines. The goal is not merely to compare individual optimizers in isolation, but rather to ask the following under the same discrete-solution task: when the objective is more closely aligned with the final output form, what search mechanism is more effective at locating high-quality parameter regions under a limited budget? The results in Sections~\ref{subsec:shot_allocation}--\ref{subsec:noise_results} show that this choice is well matched to the maximum-probability-bitstring objective.

\subsection{Shot Allocation: From Fixed Shots to Adaptive Shots}
\label{subsec:shot_allocation}

If the objective function and optimization strategy determine where to search, then shot allocation determines how much measurement at a given parameter point is sufficient. Although the fixed-shot strategy is simple to implement, it implicitly assumes that every parameter point deserves the same high-precision evaluation cost. In contrast, our approach first performs a small-shot pilot evaluation at each candidate parameter point, and then judges whether the current estimate is sufficiently stable using two criteria: mode confidence and normalized cut variance. Only when both criteria are satisfied is the current evaluation accepted; otherwise, the number of shots is increased progressively until the conditions are met or the per-point cap is reached. In this way, the measurement budget is no longer distributed uniformly, but is instead concentrated on the parameter regions that genuinely require further discrimination.

\begin{figure}[H]
    \centering
    \includegraphics[width=0.98\columnwidth]{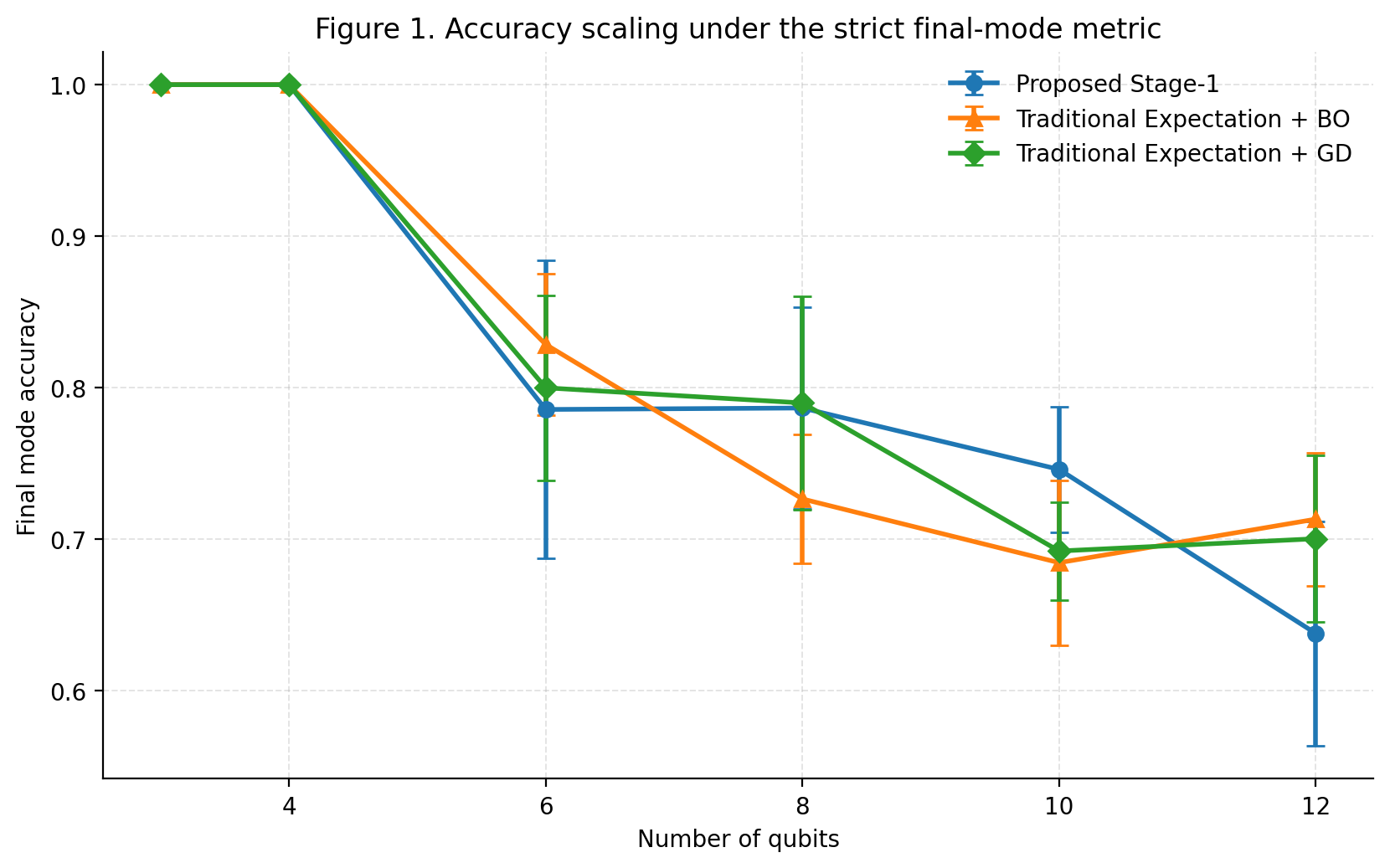}
    \caption{Comparison of the shots required to reach a final mode accuracy of $0.80$. This figure directly reflects the quantum sampling cost needed to achieve the same discrete-solution quality and is therefore taken as the main figure for the resource-efficiency analysis in this chapter.}
    \label{fig:shots_to_threshold}
\end{figure}

Figure~\ref{fig:shots_to_threshold} presents a direct comparison of the number of shots required to reach a final mode accuracy of $0.80$. This is also the central figure of this chapter for analyzing resource efficiency. In most qubit settings, the proposed method requires fewer shots than conventional expectation with Bayesian optimization to reach the same accuracy threshold, with the advantage being particularly pronounced at $n=3,4,8,10$. Even where this advantage becomes weaker at a few individual points, the overall trend remains in a lower-resource regime. In other words, adaptive allocation does not merely reduce the number of measurements; it uses the quantum sampling budget more efficiently while maintaining the same target accuracy.

\begin{figure}[H]
    \centering
    \includegraphics[width=0.98\columnwidth]{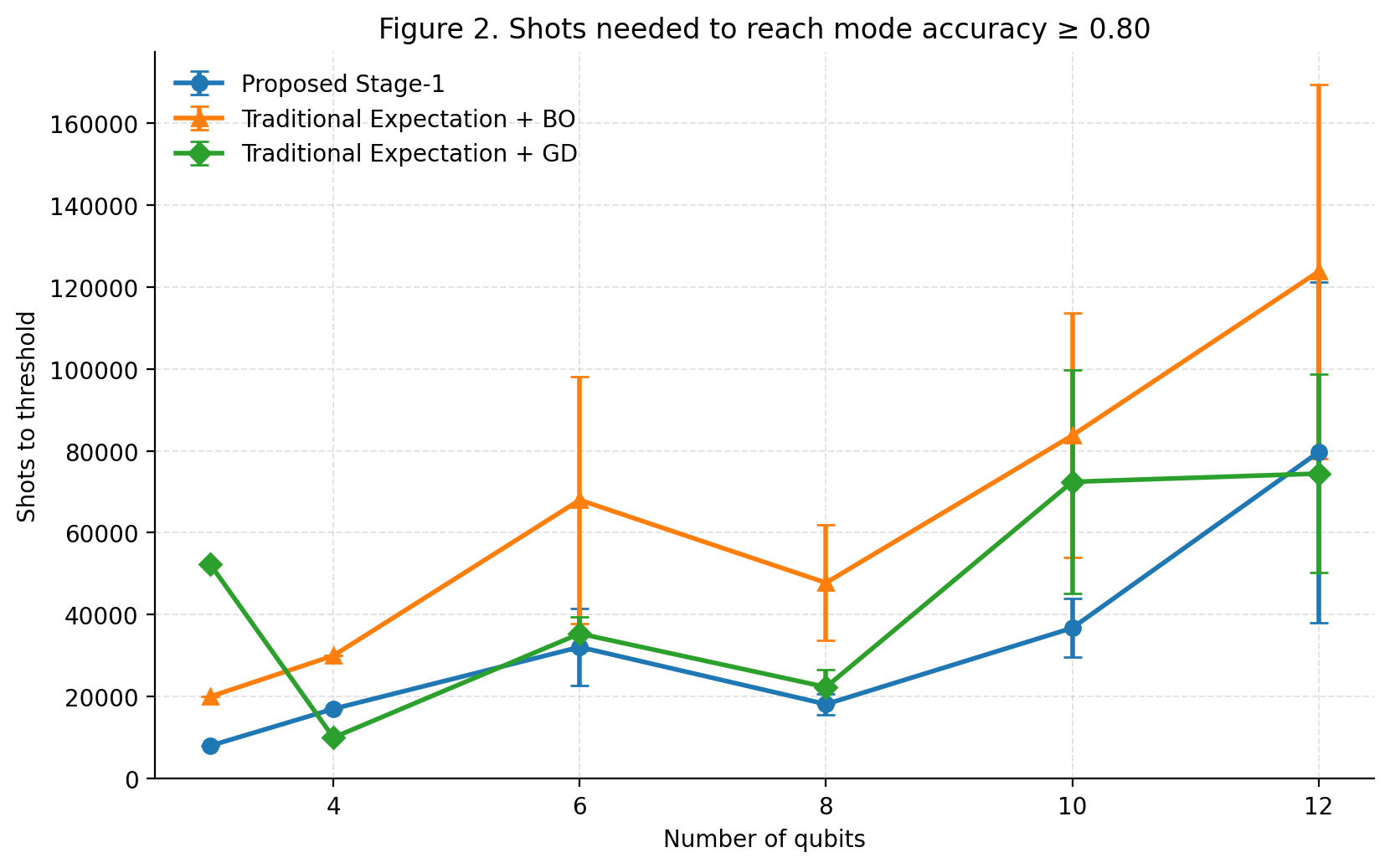}
    \caption{Resource saving rate relative to conventional expectation with Bayesian optimization under the constraint that the final mode accuracy is no lower than $0.80$. Positive values indicate that the proposed method uses fewer shots to achieve the same accuracy.}
    \label{fig:saving_rate}
\end{figure}

A comparison of absolute shot counts alone is still insufficient to fully characterize the resource savings. Therefore, Fig.~\ref{fig:saving_rate} further reports the saving rate relative to conventional expectation with Bayesian optimization. The proposed method achieves a positive average saving rate at multiple qubit sizes, indicating that fewer measurements are indeed sufficient to attain the same accuracy at those points. At a small number of sizes, fluctuations in the saving rate reflect the fact that, in a high-dimensional discrete landscape, instance-to-instance variability still affects the time required to reach the threshold. Because of this statistical variability, we do not treat the saving rate at any single point as the sole evidence, but instead interpret it together with the broader accuracy--resource trade-off.

\begin{figure}[H]
    \centering
    \includegraphics[width=0.98\columnwidth]{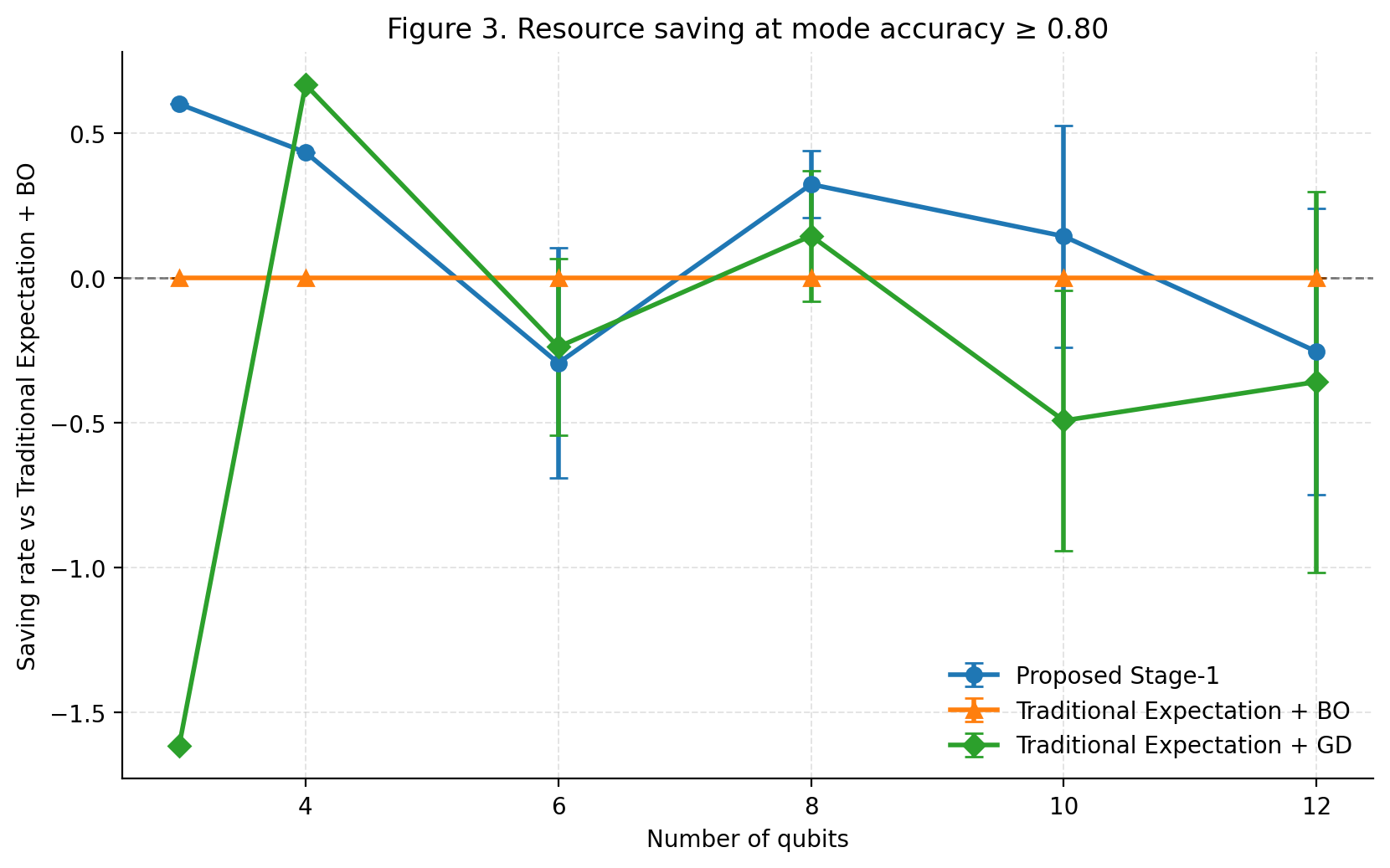}
    \caption{Accuracy--resource Pareto comparison in the qubit-scaling experiment. The horizontal axis denotes the total number of shots, and the vertical axis denotes the final mode accuracy.}
    \label{fig:pareto}
\end{figure}

Figure~\ref{fig:pareto} presents the Pareto comparison between final mode accuracy and total shots. The proposed method is generally located in a lower-shot region and extends the frontier toward the low-resource side without significantly sacrificing the final discrete-solution quality. This is consistent with the equal-accuracy comparison in Fig.~\ref{fig:shots_to_threshold}, indicating that the resource advantage does not arise from a simple compromise in solution quality, but from the joint improvement brought by the objective design, the optimization strategy, and adaptive measurement allocation.

\subsection{Overall Results from the Joint Design of the Three Components}
\label{subsec:overall_results}

\begin{figure}[H]
    \centering
    \includegraphics[width=0.98\columnwidth]{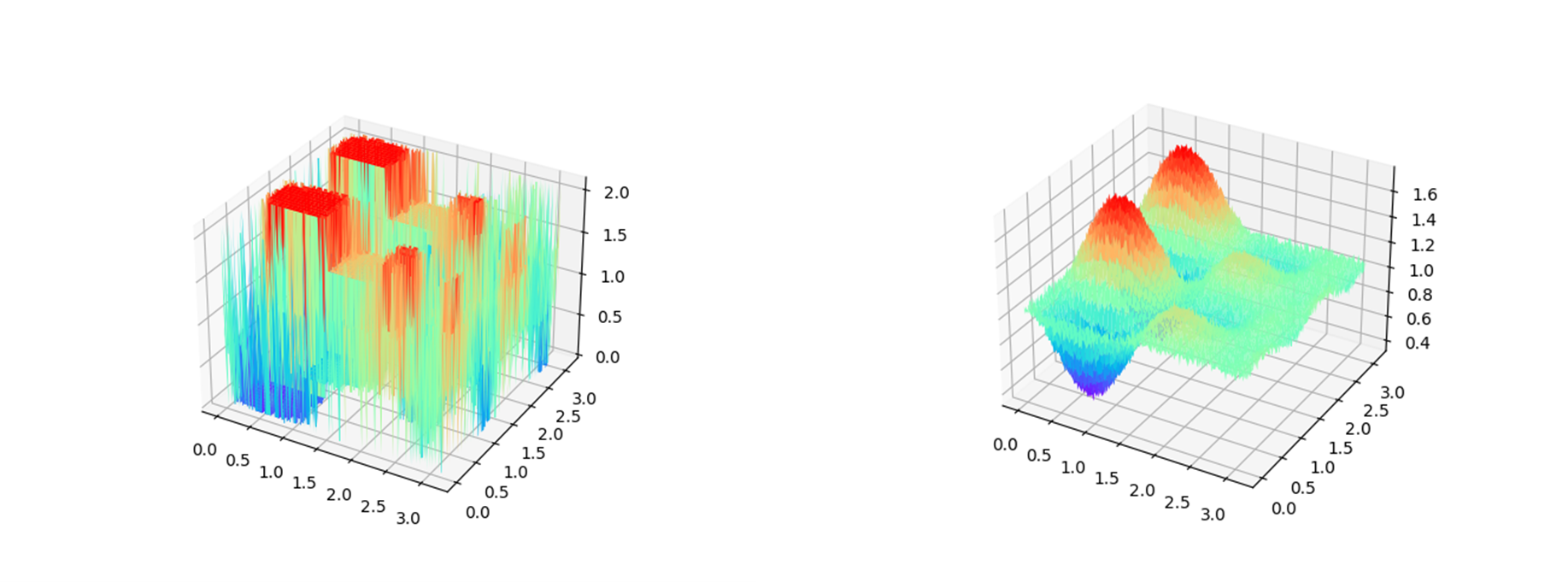}
    \caption{Final mode accuracy at different qubit scales. The blue, orange, and green curves correspond to the proposed method, conventional expectation with Bayesian optimization, and conventional expectation with gradient descent, respectively.}
    \label{fig:qubit_final_mode}
\end{figure}

Combining the designs of the objective function, the optimization strategy, and shot allocation yields the complete methodological framework of this work. Figure~\ref{fig:qubit_final_mode} shows the qubit-scaling results under the stringent final mode accuracy metric. As the problem size increases, the accuracy of all three methods declines, which is consistent with the increased difficulty caused by the rapid expansion of the discrete solution space. Nevertheless, at most intermediate scales, the proposed method still maintains final discrete-solution quality comparable to, and sometimes better than, the two baselines, while consuming substantially fewer total shots than the fixed-high-cost route based on conventional expectation with Bayesian optimization. Together with the equal-accuracy analysis in the previous subsection, this shows that the advantage does not come from relaxing the evaluation criterion, but from a higher level of consistency among the objective definition, the outer-loop search mechanism, and the resource-allocation strategy.

\begin{figure}[H]
    \centering
    \includegraphics[width=0.98\columnwidth]{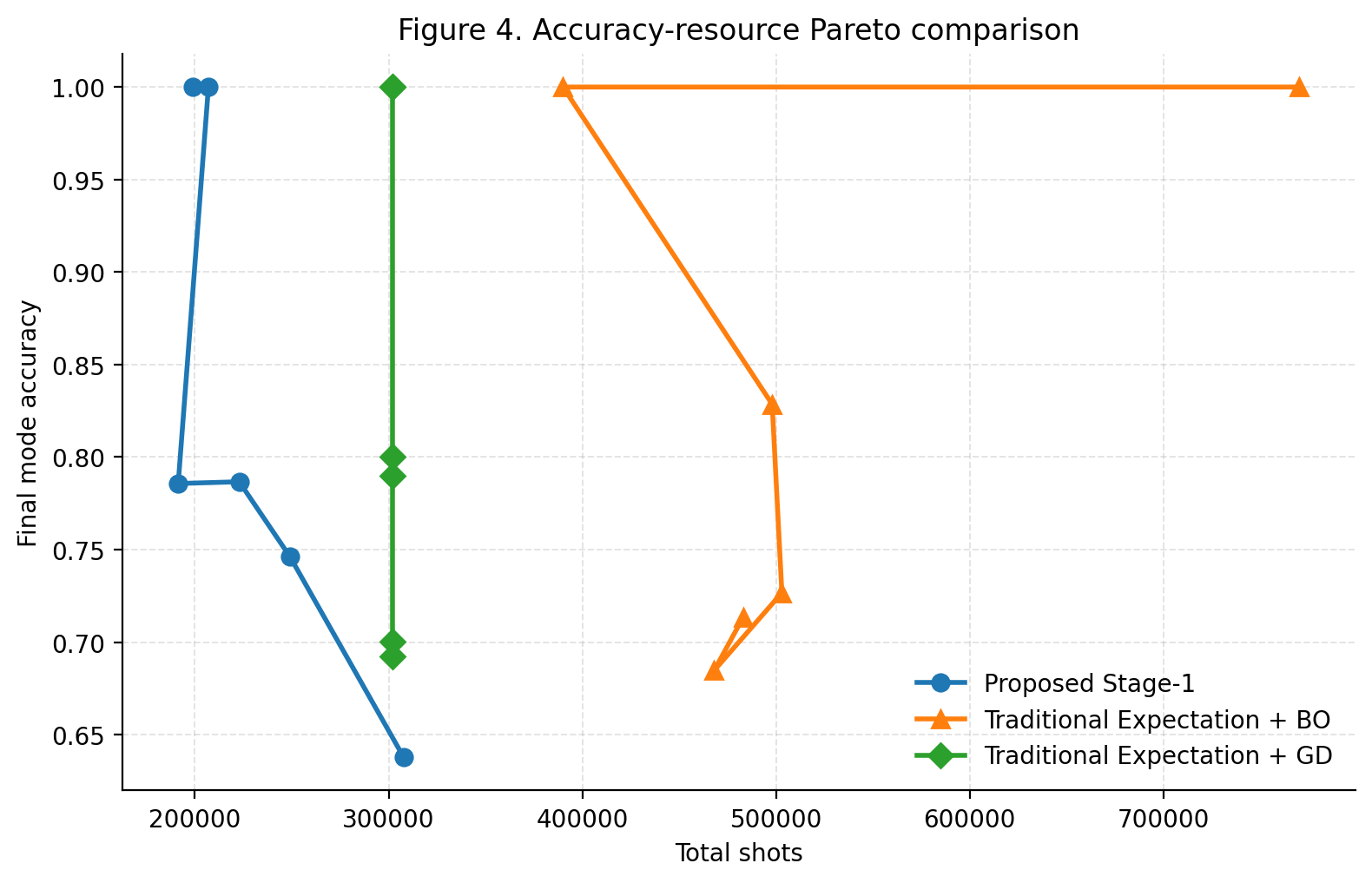}
    \caption{Results of the circuit-depth-scaling experiment. The left panel shows the final mode accuracy, and the right panel shows the corresponding total number of shots.}
    \label{fig:depth_dualpanel}
\end{figure}

The depth-scaling results provide another important perspective. The left panel of Fig.~\ref{fig:depth_dualpanel} shows that, as $p$ increases from $1$ to $6$, the final mode accuracy does not improve monotonically. Instead, the best results are concentrated at intermediate depths, with the proposed method achieving the highest average accuracy of this experiment at $p=3$. The right panel shows that, over the same depth range, conventional expectation with Bayesian optimization consistently requires a higher total number of shots, whereas the proposed method consumes markedly fewer resources at most depths. As the depth continues to increase, this advantage does not disappear; instead, it remains relatively stable due to the effect of adaptive allocation. This indicates that simply increasing circuit depth does not guarantee sustained improvement in the final output. In higher-dimensional parameter spaces, the importance of objective design and resource-allocation mechanisms becomes even more pronounced.

Overall, the improvement delivered by the proposed method does not arise from a single technique, but from the coordination of three design layers. Centering the objective function on the final dominant bitstring ensures that the evaluation criterion is aligned with the final output; Bayesian optimization reduces the dependence on local smoothness; and adaptive shot allocation further suppresses unnecessary measurements. Together, these components yield a more favorable accuracy--resource trade-off under a finite sampling budget.

After Stage~1 has identified a promising parameter region, Stage~2 introduces a lightweight post-processing step aimed at amplifying the probability of the target bitstring. Its role is not to replace the main optimization pipeline, but to further increase the dominance probability of the target bitstring once a high-quality discrete solution has already been obtained. Because this stage operates only within a local parameter neighborhood, its additional cost remains relatively controllable, making it better understood as an engineering enhancement module.

In the current statistics, the overall gain from Stage~2 does not constitute the central conclusion of this work. We therefore do not discuss it extensively, and keep the emphasis on the main framework represented by Stage~1. In general, Stage~2 is better regarded as an optional enhancement on top of an already good solution. When the resource budget allows and a further increase in the observable probability of the target bitstring is desired, it can provide a useful supplement; however, it does not alter the main conclusion above regarding the synergy among objective design, optimization strategy, and adaptive shot allocation.

\subsection{Noise Analysis}
\label{subsec:noise_results}

\begin{figure}[H]
    \centering
    \includegraphics[width=0.98\columnwidth]{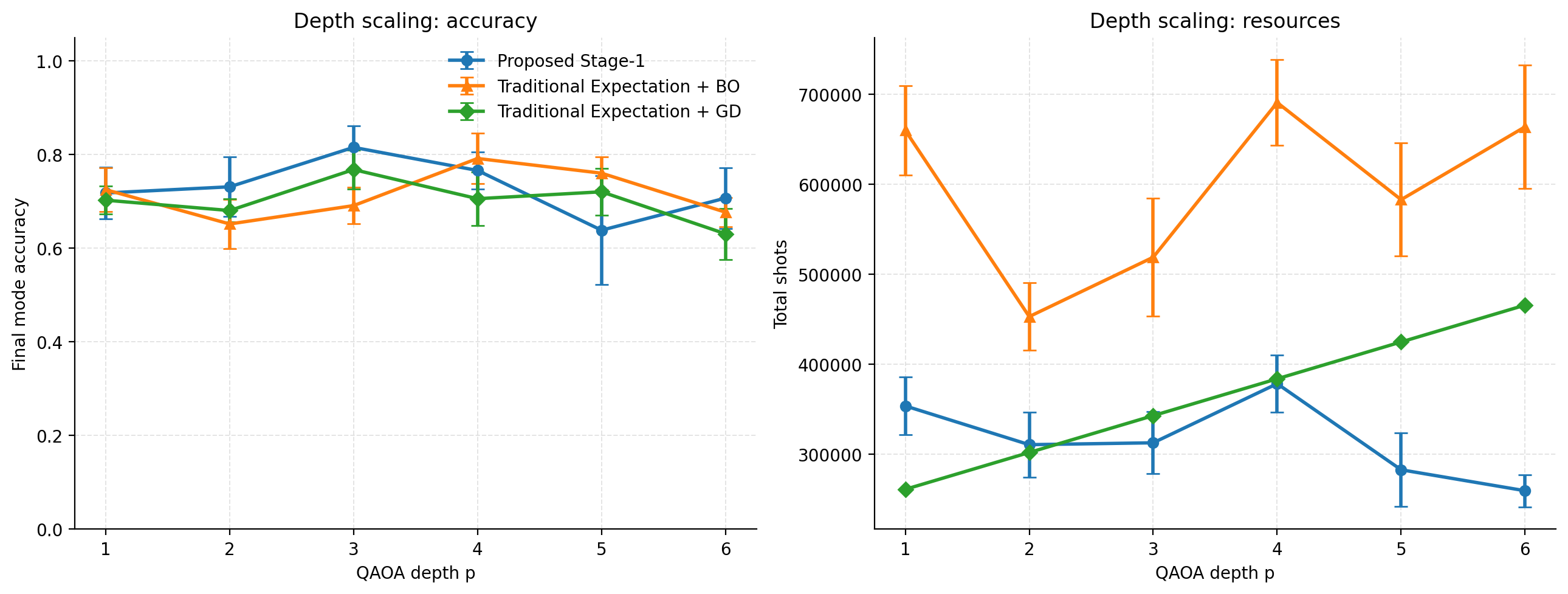}
    \caption{Accuracy--resource Pareto comparison under the depolarizing-noise sweep. The horizontal axis denotes the total number of shots, and the vertical axis denotes the final mode accuracy.}
    \label{fig:noise_pareto}
\end{figure}

To evaluate the method under conditions closer to practice, we further introduce a depolarizing-noise sweep and compare the accuracy and resource consumption of different methods with $n=10$ and $p=2$ fixed. Figure~\ref{fig:noise_pareto} shows the accuracy--resource Pareto relation under noise. The proposed method remains in a lower-resource regime at most noise levels, while its corresponding final mode accuracy is comparable to, and in some cases higher than, those of the two baselines. By contrast, conventional expectation with Bayesian optimization is distributed more broadly in a higher-shot region, whereas conventional expectation with gradient descent can achieve relatively high accuracy at a few individual noise points but exhibits larger overall fluctuations. These results indicate that, even after the introduction of additional statistical disturbances by noise, the proposed method is still able to maintain competitive accuracy.

\begin{figure}[H]
    \centering
    \includegraphics[width=0.98\columnwidth]{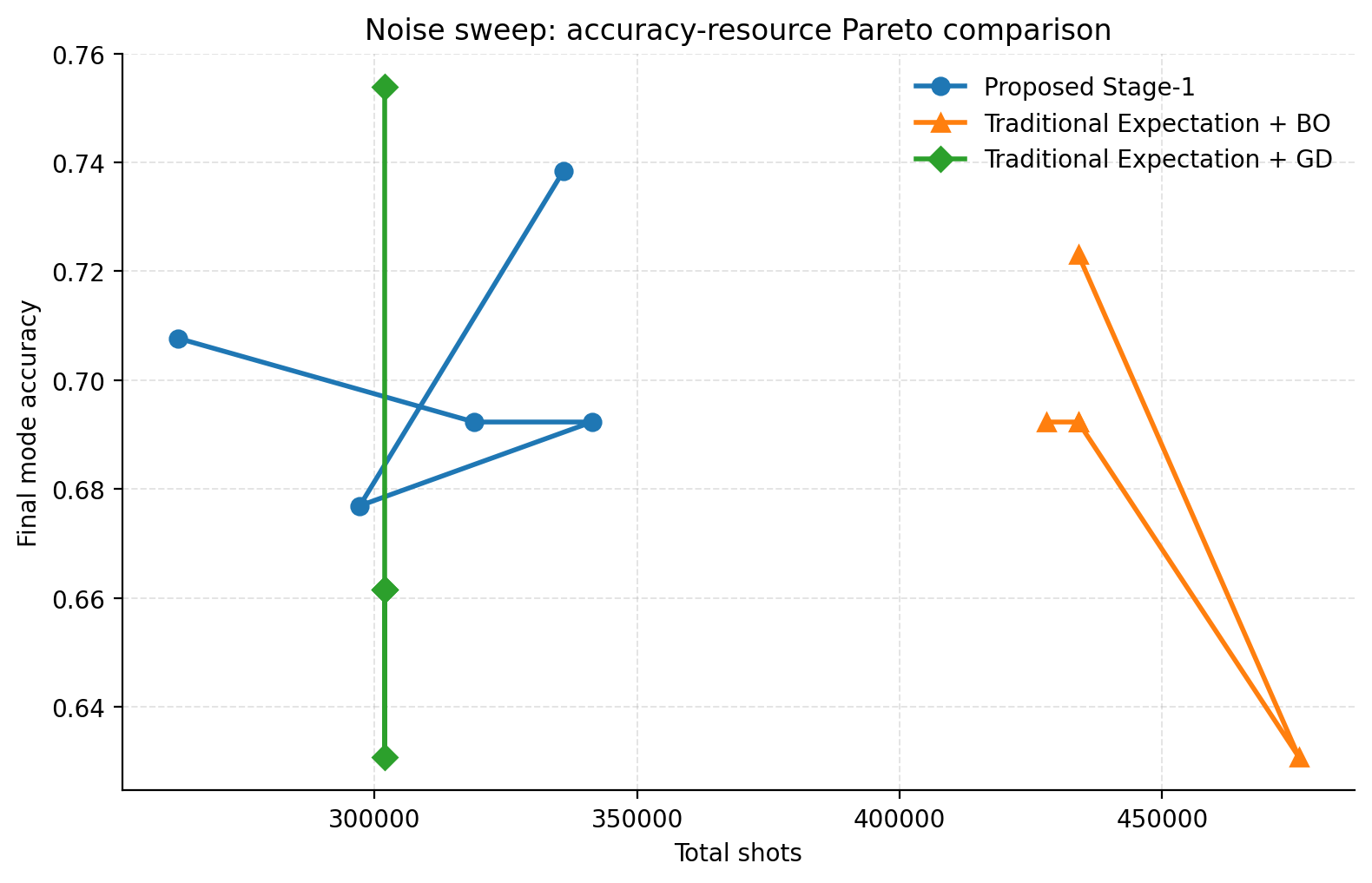}
    \caption{Auxiliary results for the noise sweep. The left panel shows the variation of final mode accuracy with depolarizing-noise strength, and the right panel shows the variation of the total number of shots with noise strength.}
    \label{fig:noise_aux}
\end{figure}

Figure~\ref{fig:noise_aux} further provides a two-panel view of how accuracy and total resource consumption vary with noise strength. The left panel shows that, within the noise range considered here, the final mode accuracy of the proposed method remains broadly stable and does not exhibit a systematic collapse as the noise increases. The right panel shows that conventional expectation with Bayesian optimization consistently incurs the highest shot cost throughout the entire noise sweep, whereas the proposed method remains stably in a lower-resource regime. This indicates that, even in the presence of noise, the pilot-evaluation-based adaptive allocation can still effectively suppress unnecessary measurement overhead, rather than degenerating into an evaluation strategy as expensive as the fixed-shot route.

\section{Discussion and Conclusion}
\label{subsec:discussion_conclusion}

This work presents a resource-efficient QAOA framework for obtaining high-quality discrete solutions under a limited sampling budget. Instead of treating the objective function, the outer-loop optimizer, and the measurement budget as independent design choices, we combine them in a single co-design strategy. At the objective level, we replace the conventional expectation value by the cut value of the most probable measured bitstring, so that the optimization target is more directly aligned with the final output of a combinatorial optimization task. At the search level, we employ Bayesian optimization to cope with the discrete, noisy, and piecewise-flat landscape induced by the mode-based objective. At the resource-allocation level, we use a dual criterion based on mode confidence and normalized cut-value variance, so that additional shots are spent only when statistical uncertainty still warrants further discrimination. Together, these components improve the consistency between evaluation, search, and resource use.

The numerical results support three main observations. First, on 3-regular MaxCut, the mode-based objective achieves discrete-solution quality comparable to that of the conventional expectation-based objective in both unweighted and weighted settings, indicating that the new objective is not merely a crude surrogate but a practically meaningful task-aligned alternative. Second, when the final mode accuracy threshold is fixed, the proposed framework typically requires fewer total shots, and it occupies a more favorable region in the accuracy--resource Pareto comparison. This indicates that the observed resource advantage does not arise from relaxing the evaluation criterion, but from the joint effect of objective redesign, Bayesian search, and adaptive sampling. Third, in the depth-scaling study, the performance does not improve monotonically with circuit depth; instead, the best results appear at intermediate depths, which suggests that, under finite resources and realistic noise, increasing ansatz depth alone is not the most effective route to better practical performance.

At the same time, the present results should be interpreted with appropriate caution. The proposed framework is better understood as a method for lowering the practical resource threshold of QAOA than as a proof of a strict complexity-theoretic quantum advantage. Our experiments are restricted to small- and medium-scale 3-regular MaxCut instances, and several hyperparameters in the adaptive allocation and Bayesian search remain empirically chosen. In addition, although the mode-based objective is better aligned with the final discrete output, it also induces a more discontinuous landscape and may become less stable when several high-probability bitstrings are nearly degenerate. These features do not invalidate the method, but they do indicate that the present conclusions are primarily methodological and empirical rather than asymptotic.

Several directions remain open for future work. It will be important to test the framework on broader classes of combinatorial optimization problems and graph ensembles, to study its behavior under hardware-calibrated noise models and on real devices, and to develop a more systematic theory for the statistical properties of the mode-based objective and the convergence behavior of adaptive-shot optimization. Overall, the central message of this work is that, for QAOA applied to discrete optimization, practical performance can be improved not only by modifying the ansatz, but also by redesigning what is optimized, how it is searched, and where the measurement budget is spent. This perspective provides a useful route toward more effective quantum optimization under realistic resource constraints.

\section{Acknowledgements}
This work is sponsored by the CPS-Yangtze Delta Region Industrial Innovation Center of Quantum and Information Technology-MindSpore Quantum Open Fund. All numerical simulations presented in this work were implemented using the MindSpore Quantum framework\cite{xu2024mindspore}, which provides a powerful environment for quantum computation simulation. The source code and data required to reproduce the numerical results reported in this paper are publicly available at https://github.com/Siranopenaccess/Resource-Efficient-QAOA-via-Bayesian-Optimization-and-Maximum-Probability-Bitstring-Evaluation.

    \bibliographystyle{apsrev}
    \bibliography{note.bib}

\end{document}